\documentclass[conference]{IEEEtran}

\usepackage{blindtext}
\usepackage{eso-pic}
\title{Sample paper}

\author{\IEEEauthorblockN{author 1\\ Department 1}%
  \and
  \IEEEauthorblockN{author 2\\ Department 2}
  \IEEEauthorblockA{Institution/university \\
    mail1,mail2,mail3,mail4} \and
  \IEEEauthorblockN{author 3\\ Department 3}
}

\usepackage{eso-pic}
\newcommand\AtPageUpperMyright[1]{\AtPageUpperLeft{%
 \put(\LenToUnit{0.5\paperwidth},\LenToUnit{-1cm}){%
     \parbox{0.5\textwidth}{\raggedleft\fontsize{9}{11}\selectfont #1}}%
 }}%
\newcommand{\conf}[1]{%
\AddToShipoutPictureBG*{%
\AtPageUpperMyright{#1}
}
}

\IEEEoverridecommandlockouts
\usepackage{cite}
\newcommand{\ts}{\textsuperscript}
\usepackage{amsmath,amssymb,amsfonts}
\usepackage{algorithmic}
\usepackage{graphicx}
\usepackage{textcomp}
\usepackage{xcolor}
\usepackage{fancyhdr}
\usepackage{url}

\makeatletter
\def\ps@IEEEtitlepagestyle{%
  \def\@oddfoot{\mycopyrightnotice}%
  \def\@evenfoot{}%
}
\def\mycopyrightnotice{%
  {\footnotesize 978-1-7281-7116-6/20/\$31.00 ©2020 IEEE \hfill}
  \gdef\mycopyrightnotice{}
}

\def\BibTeX{{\rm B\kern-.05em{\sc i\kern-.025em b}\kern-.08em
    T\kern-.1667em\lower.7ex\hbox{E}\kern-.125emX}}

\begin{document}


\title{Textual analysis of End User License Agreement for red-flagging potentially malicious software\\

}

\author{\IEEEauthorblockN{ Behraj Khan}
\IEEEauthorblockA{\textit{dept.(of Computer Science )} \\
\textit{FAST-NU}\\
Karachi, Pakistan \\
behraj.khan@nu.edu.pk}
\and
\IEEEauthorblockN{Tahir Syed}
\IEEEauthorblockA{\textit{dept.(of Computer Science )} \\
\textit{FAST-NU}\\
Karachi, Pakistan \\
tahir.syed@nu.edu.pk}
\and
\IEEEauthorblockN{Zeshan Khan}
\IEEEauthorblockA{\textit{dept. (of Computer Science )} \\
\textit{FAST-NU}\\
Karchi, Pakistan }
\and
\IEEEauthorblockN{Muhammad Rafi}
\IEEEauthorblockA{\textit{dept. (of Computer Science )} \\
\textit{FAST-NU}\\
Karachi, Pakistan }
}

\maketitle
\conf{Proc. of the 2\ts{nd} International Conference on Electrical, Communication and Computer Engineering (ICECCE)\\ 
14-15 April 2020, Istanbul, Turkey}

\begin{abstract}
New software and updates are downloaded by end users every day. Each dowloaded software has associated with it an End Users License Agreements (EULA), but this is rarely read. An EULA includes information to avoid legal repercussions. However,this proposes a host of potential problems such as spyware or producing an unwanted affect in the target system.  End users do not read these EULA's because of length of the document and users find it extremely difficult to understand. Text summarization is one of the relevant solution to these kind of problems. This require a solution which can summarize the EULA and classify the EULA as "Benign" or "Malicious". We propose a solution in which we have summarize the EULA and classify the EULA as "Benign" or "Malicious". We extract EULA text of different sofware's then we classify the text using eight different supervised classifiers.  we use ensemble learning to classify the EULA as benign or malicious using five different text summarization methods. An accuracy of $95.8$\% shows the effectiveness of the presented approach.

\end{abstract}

\section{Introduction}


Spywares aims to gather information about a person without their consent or permission. In early stages of spywares, these software's just gather private information but now they can alter the information and can degrade user performance by changing configurations of the systems. According to a report about McAfee, it is stated "only about one in ten internet users say the current practice of clicking through a user agreement or disclaimer is adequate consent to install adware on a person's computer".
Due to growth of textual information on the internet and the access to this information by users through different types of devices, it is necessary that the users can access the important sentences of information in summary form, which will generally describe the main idea. 
The need of document summarization is growing rapidly because the sizes of the documents are increasing at an unprecedented rate. Many challenges exist in this domain, some of them are: 

\begin{itemize}
	\item Sentence importance: \\
Important sentences, which are used more often, should be considered in summary and least frequent words should be neglected.
\item Semantic correctness:\\
The words should make sense, the whole document should give the summary on behalf of sentence.
\item Grammatical correctness:\\
Sentences used in the summary should not contain any type of grammar mistake.\\
\end{itemize}
The main idea of this research is that spyware can violate your rights by gathering information and revealing it without your explicit permission. So, users have to be aware of these spyware threats. It can be achieved by a system that classify License agreements of the systems before installation and notify users by the nature of that EULA and then they can continue the installation of the software if they want.\\

When we talk about classification of EULA, EULA are classified into many categories, but the main classification will be whether the EULA is legitimate (benign) or associated with spyware (malicious) code and risk someone's system. Malicious EULA’s are those which hosts spyware but due to difficult terminologies and the amount of text presented in EULA, it cannot be identified easily. The increase of spyware has resulted in vast numbers of users experiencing loss of control over personal information and decreased computer performance and someone system crashes.\\

Document summarization has become very important since the exponential growth of documents, and in our case specially the growth of legal documents. As the data is growing, it is very much needed that there must be some summarization methods which will serve the purpose of replacing a larger document with a smaller summary so that readers only find the information they need.
 Motivating form  this problem we propose to summarize the EULA for clarification.\\

The user download new software application and start installation when needed. The vendor of the downloaded application may ask the user for some permission the user is not aware of and can spy on that user, or worse. Text summarization gives us the advantages:
\begin{itemize}
  \item Create summary of a document.
	\item Shortening the length of the document
	\item Highlighting most used words, sentences.
\end{itemize}

We propose a method which classify EULAs on the basis of a binary benign/malicious classification. The extracted EULA text is classified using the EULA dataset\cite{IEEEHow:Lay}  using ensemble learning of eight supervised machine learning classifiers. The purpose of this research is to analyze a legal document and create a summary for that document. In this research we consider EULA (End User License Agreement) which is a legal document for software.\\

our algorithm reduces the length of the document by using techniques proposed by the text summarization researchers. In this research we focused on single document and extractive document summarization. Single document means that the research will focus on a EULA present in a single document, and will not analyse multiple document. Extraction based summarization is to use the words which are present in the original document without modifying the words. There are still many problems associated with single document summarization. The research only focuses on single document and the extraction of document algorithms for the summarization of the EULA.\\

The rest of the paper is organized as follows: Section II explains the problem, gives its context and mentions related strategies. Section III gives the bird's-eye-view of proposed methodology. Section IV compares the performance of the proposed strategy and motivates the differences. Section V concludes the paper.

\section{Related Work}

Automatic summarization system is defined as a system designed to take a single document (single document summarization) or multiple documents (multiple document summarization) containing a large amount of text and produce a concise and fluent summary. 
Extractive summaries or extracts are produced by concatenating the important sentences chosen from the materials provided to the summarization system whereas abstractive summaries (abstracts) are written to communicate the main idea or information in the input and may reprocess phrases or clauses from it, but the summaries are in general articulated in the words of the text under consideration.\\

Summaries are differentiated by their content as well. An indicative summary tells the user specific information about the summary generated. It may provide characteristics such as length and writing style.   An informative summary on the other hand will consist of facts that are reported in the input document(s).

Similarly, the purpose of summarization can be to produce a generic summary of the document, or to summarize the content that is most relevant to a user query[4]

Text classification being used for spam filtering, news filtering, sentiment analysis and many more.
In 2010, Niklas Lavesson et al. \cite{IEEEHow:Lay} compare the performance of 17 text classifiers by classifying EULA into two classes legitimate(benign) or spyware(malicious). Learning algorithms that learned from the pre-defined dataset are known as supervised machine learning algorithms.\\
(ToS;DR) rate the terms of service and policies of major internet sites and services by analyzing. Each aspect of a TOS or privacy policy is assessed as positive, negative, or neutral. Services are graded from A (best) to E (worst) once project associates form an agreement on it.
There had been a massive research present in text summarization and there are many methods to generate a summary out of the text. But there is still a gap in summarization of license agreements. Summarization can be achieved by taking out the most important sentences from a document and arrange it in a readable style. Using different techniques, a reduced version of document can be obtained.\\

TextRank is a graph based model which extracts sentences from the document. Since it is independent of language as it does not require linguistic knowledge and domain, it is widely used for Extractive Summarization. In the year 2016, there was a research that claims to improve TextRank by accommodating it with different text retrieval and indexing functions. Given that TextRank performs 2.84\% over the baseline, our improvement of 2.92\% over the TextRank score is an important result [4]. The BM25-TextRank algorithm was also contributed to the Gensim project. This is a modern approach that fits quite nicely with our desired approach. Hence, this was the algorithm chosen in order to effectively summarize EULA.\\

TLDRLegal is a website launched in 2012. Its main purpose is to aware people that do not read terms and conditions because they were too long. So, they process each license agreement through the proper channel of renowned lawyers and made a short, plain English summary of that EULA. We used this website as a reference to check the automated summary produced by Gensim as we lack summarized dataset of EULA.\
The focus of this paper and subsequent prototype is to propose an extractive single-document summarization technique that can be applied effectively for summarization and review the contents of the EULA and classify the EULA either as benign or malicious on the basis of results taken from eight classifiers.\
The amount of spyware has increased dramatically due to the high value for marketing companies of the information that is collected. Spyware is designed to collect user information for marketing campaigns without the informed consent of the user. This type of software is commonly spread by bundling it with popular applications available for free download. [3]\\

The very first work done on automatic summarization was done by Luhn. The idea was simple that some words in the document are expressive of its content. He also suggested using the frequency of words in order to determine this as the words that occur repeatedly are most likely to contain the main topic or idea that is conveyed in the document. He also gave the notion of using and defining a stop word list. A stop word list is a list that contains pronouns, determiners and prepositions that add meaning to the natural language however they are of little or no use to the automatic summarization systems. \\

The summaries must be 20 percent or one-fifth of the actual material provided \cite{IEEEhowto:nen} and must fulfill the criteria of being able to produce and focus on the main idea conveyed, must be grammatically correct and correctly ordered. Furthermore, the summary to be produced is to be extractive. 

Once a correct summary is generated then text to text generation techniques can be applied. One of the other important focal point of this project is to produce an informative summary [4]. The automatically generated summary must not suffer from exophora, Paice [2, 3] Exophora is defined as having unresolved references to sentences in the given document but is not incorporated into the summary.

\section{Proposed System}
The data have been fetched from tldrlegal.com a website that contains software licenses in plain English. Then five types of different text summarization algorithms have been applied on the dataset for summarizing the each EULA(End User License Agreement). The summarized EULA is then provided as input to different type of text classifier for identification of malicious/good software. The decision about a particular summarized EULA that wether it's malicious or good based on majority vote result by using ensembling on these classifiers.following is the algorithm for above mentioned task and for the system architecture mentioned in \ref{fig:sad}\newline

\textbf {Proposed Algorithm:}

corpus containing list of EULA's:
\[ \left \{X = X_1, X_2, X_3,...,X_n\right \} \] 
Class of document in the class:
\[ \left \{Y = Y_1, Y_2, Y_3,...,Y_n\right \} \]
the license agreement and the sentences of that agreement
\[ \left \{x_i = s_1, s_2, s_3,...,s_n\right \} \]
training set
\[ \left \{X_t = X_1, X_2, X_3,...,X_n\right \} \] 
list of classifiers for classifying training set
\[ \left \{C = c_1, c_2, c_3,...,c_n\right \} \] 
list of summarizer used for license agreements
\[ \left \{S_m = s_1, s_2,s_3,s_4,s_5\right \} \] 

\[
\text{training}\left(X,C,S_m\right):
\]

\[ \left \{X_{tr},X_{ts},Y_{tr},Y_{ts} = \text{split\_train\_test}\left(X,Y\right)\right \} \]

\text{maxacc} = \text{maxsmry} = 0

\[\text{for each } c \in C: \]
\[\hspace{1cm} c.\text{train}\left(X_{tr}, Y_{tr}\right) \]
\[\hspace{1cm} \text{acc} = c.\text{validate}\left(X_{ts}, Y_{ts}\right) \]
\[\hspace{1cm} \text{maxacc} = \max\left(\text{acc},\text{maxacc}\right) \]
\[\hspace{1cm} \text{classifier} = c\]

\[\text{for each } sm \in S_m: \]
\[\hspace{1cm} sm.\text{train}\left(X_{tr}, Y_{tr}\right) \]
\[\hspace{1cm} \text{smry} = sm.\text{validate}\left(X_{ts}, Y_{ts}\right) \]
\[\hspace{1cm} \text{maxsmry} = \max\left(\text{smry},\text{maxsmry}\right) \]
\[\hspace{1cm} \text{summarizer} = sm\]

\text{return classifier,summarizer}

\[\text{Test}\left(X_n\right): \]
\[\hspace{1cm} \text{classifier.test}\left(X_n\right) \]
\[\hspace{1cm} \text{summarizer.test}\left(X_n\right)\]

\subsection{Document Summarization}
The document summarization algorithms which we encountered are listed below:
\subsubsection{LexRank}
It is a graph based model. Vertices represent sentences and edges between sentences are assigned weights equal to the similarity between the two sentences. The method most often used to compute similarity is cosine similarity with TF-IDF weights for words. Graph representation is more flexible than sentence clustering. Vertex importance or centrality can be computed using general graph algorithms, such as PageRank.

\subsubsection{Random Indexing Approach}
Random Indexing for improvements both in terms of formation of index vectors of the document, and construction of context vectors by using convolution vectors. As a consequence, three improved versions of the algorithm, viz. RISUM, RISUM+ and MRISUM were obtained. The results are quite satisfactory.
\subsubsection{	Latent Semantic Analysis}
 
It is an algebraic-statistical method which extracts hidden semantic structures of words and sentences. It is an unsupervised approach which does not need any training or external knowledge. LSA uses context of the input document and extracts information such as which words are used together and which common words are seen in different sentences. High number of common words among sentences indicates that the sentences are semantically related.

\subsubsection{TextRank}

It is a graph-based ranking model for text processing. The units to be ranked are therefore sequences of one or more lexical units extracted from text, and these represent the vertices that are added to the text graph. The vertices added to the graph can be restricted with syntactic filters, which select only lexical units of a certain part of speech. it does not require deep linguistic knowledge, nor domain or language specific annotated corpora, which makes it highly portable to other domains, genres, or languages.\\

\subsubsection{	Bm25 (Modified version of TextRank)}
The combination of TextRank with modern Information Retrieval ranking functions such as BM25 and BM25+ creates a robust method for automatic summarization that performs better than the standard techniques used previously. Based on these results the use of BM25 along with TextRank for the task of unsupervised automatic summarization of texts is very beneficial.\\

\begin{table}[h]
\centering
\caption{Test of classification}
\label{my-label}
\begin{tabular}{|p{.75cm}|p{.5cm}|p{.75cm}|p{2.5cm}|p{2.5cm}|}
\hline
software Name & Result                     & Tosdr points & Source                                                       & Why malicious                                                                                                                               \\ \hline

Apple              & \%  100.0  & malicious          & http://www.apple.com\par/legal/internet-services/terms/site.html & Change of Terms Without Notice                                                                                                        \\ \hline
.Net framework 4.0 & \%  100.0 & benign         & https://msdn.microsoft\par .com/en-us/library/ms994405\par .aspx       & Mention of third party nothing much else                                                                                              \\ \hline
Steam              & \%  100.0 & benign         & http://store.steampo\par wered.com\par/eula/eula\_39190               & Notified 30days before change in Eula                                                                                                 \\ \hline
Moboplay           & \%  100.0  & malicious          & http://moboplay.com\par/agreement/pc.html                    & unsolicited or unauthorized advertising, promotional messages, third party software, web links, terminate of Agreement without notice \\ \hline

\end{tabular}
\end{table}
The example dataset we run on our product give us the result shown in table \ref{my-label}

\begin{figure}[h]
\centering

\includegraphics[width=9cm,height=6.5cm]{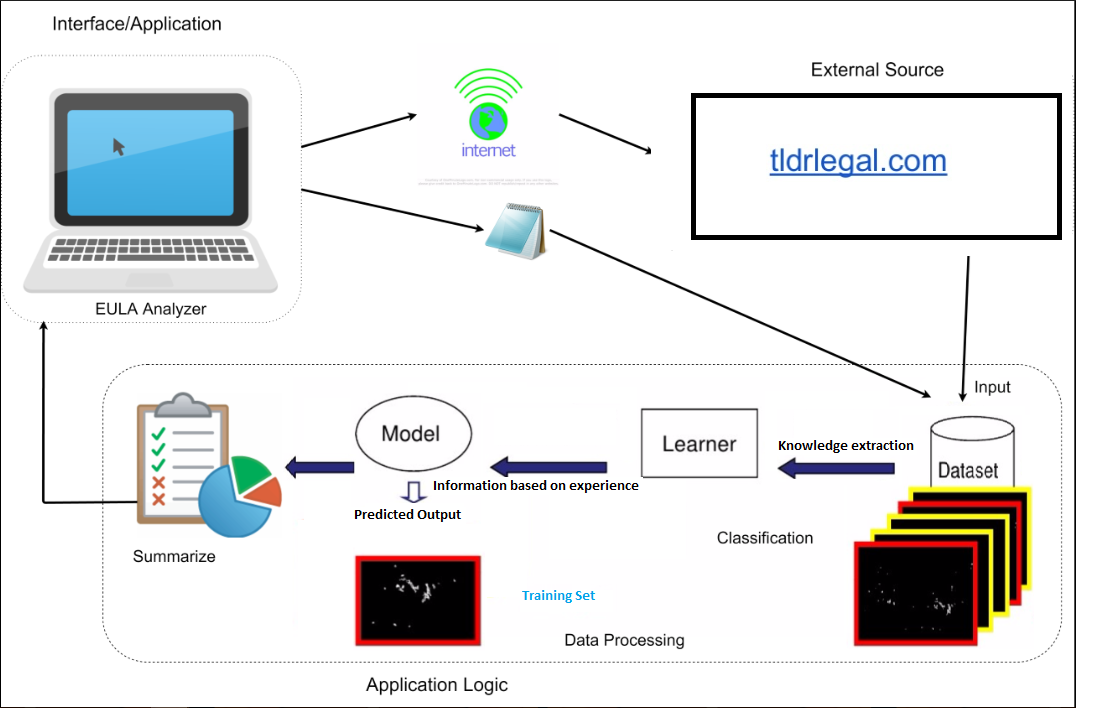}
\caption[System Architecture Diagram]{System Architecture Diagram}
\label{fig:sad}
\end{figure}
Figure \ref{fig:sad} shows the system architecture. System will work in a layered approach. The top layer will have the user interface which will have the GUI where the user will select the EULA either from open file or selecting existing EULA. Then the our application will request the data from the tldrlegal.com through internet or the user can directly open file from local storage. After that, it will go through the processing stage which includes classification and summarization. Classification uses ensemble learning to take the votes of 8 classifiers which we trained on 996 EULA’s.

(ToS;DR) rate the EULA after reading it carefully. The process of Tosdr is manual, which require human efforts. So we compared our classification result with tosdr and the result can be seen in table \ref{my-label}. 
Tldrlegal website also generate EULA summary manually, which is written by lawyers. 

\section{Experiements and Discussion}
\subsection{Datasets}
\begin{figure}[h]
\centering

\includegraphics[width=9cm,height=5.5cm]{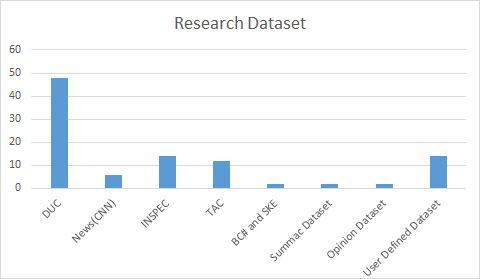}
\caption[System Architecture Diagram]{Research Dataset}
\label{fig:rd}
\end{figure}

For summarization, the distribution of studies with respect to research dataset has been indicated in figure. Statistics shows the high number 48\% of DUC dataset, 14\% of User defined dataset, 14\% of Inspec, 12\% of TAC, 6\% of newspaper, 2\% of BC3,SKE, SUMMAC, Opinoisis dataset as shown in figure \ref{fig:rd}.

We have used a pre-defined dataset comprises of 996 EULAs in which 900 are benign EULAs or contain no spyware whereas 96 EULAs are considered to be malicious.

\subsection{Classification}
The extracted EULA text is classified using the EULA dataset which consists of 996 EULAs divided into 900 benign EULAs and 96 malicious EULAs. Ensemble learning is used to take input from all eight classifiers.  The classification result of each classifier is mentioned in table \ref{my-label2}.
\begin{table}[h]
\centering
\caption{Classification Results}
\label{my-label2}
\begin{tabular}{|l|l|l|l|}
\hline
\textbf{Name Of Classifier}   & \textbf{Correctly Classified} & \textbf{Testing Set} & \textbf{Accuracy} \\ \hline
Multinomial naive bayes       & 270                           & 295                  & 91.4              \\ \hline
Bernoulli naive bayes         & 243                           & 295                  & 82.4              \\ \hline
Logistic Regression algorithm & 277                           & 295                  & 94                \\ \hline
SGD classifier                & 283                           & 295                  & 95.8              \\ \hline
SVC algorithm                 & 268                           & 295                  & 90.7              \\ \hline
Linear SVC                    & 276                           & 295                  & 93.7              \\ \hline
NuSVC algorithm               & 277                           & 295                  & 93.8              \\ \hline
\end{tabular}
\end{table}

\begin{figure}[htb]
\centering

\includegraphics[width=9cm,height=6.5cm]{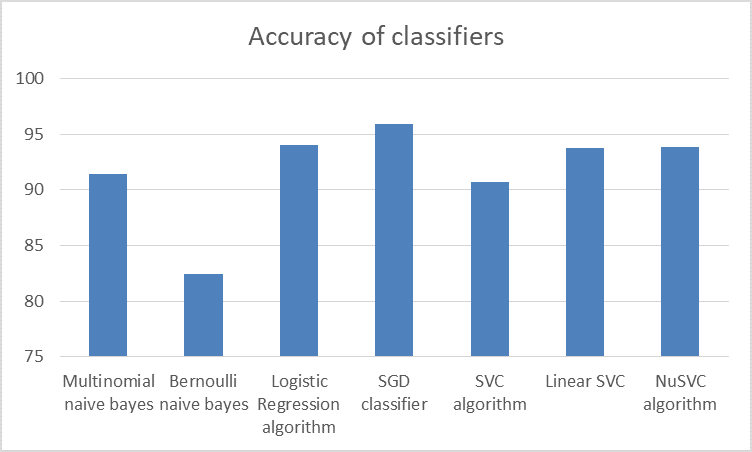}
\caption[System Architecture Diagram]{Classification result graph}
\label{fig:crg}
\end{figure}

Figure \ref{fig:crg} show the comparison results of eight classifiers which we have used for text classification. From the figure we can conclude that SGD perform very well during text classification and after that linear regression and linear SVC shows promising results.

\begin{figure}[h!]
\centering

\includegraphics[width=9.75cm,height=6.5cm]{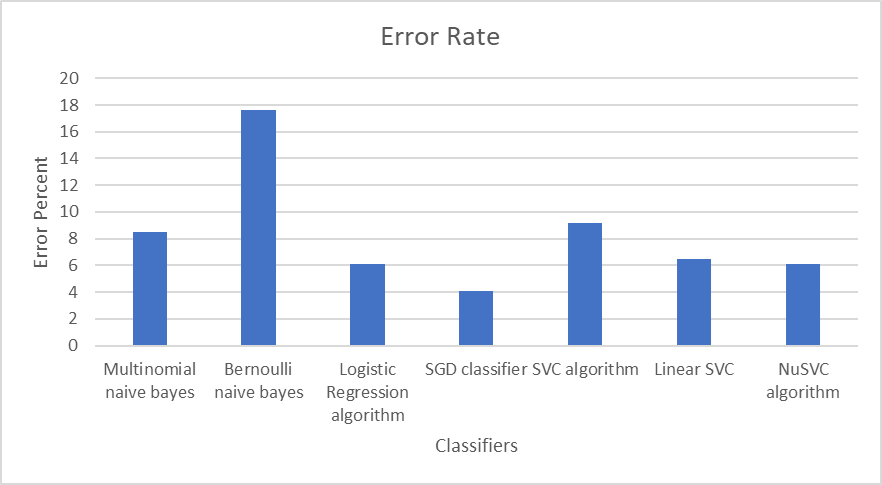}
\caption[System Architecture Diagram]{Error Rate of classifiers}
\label{fig:ErC}
\end{figure}

The figure \ref{fig:ErC} shows the comparison of error rate of the classifiers which we have used for text classification in our project. It can be concluded from the figure that the error rate of Bernouli Naïve Bayes classifier is very high. While the error rate of SGD classifier is very low, after SGD classifier Linear SVC and Logistic Regression error rate is almost same but less than Bernouli Naïve Bayse classifier.

\section{Conclusion}
In this paper we have attempted the EULA summarization and clarification problem.The dataset which we have used in this reseach contain EULA's of different online web applications. The proposed study of this paper has implemented various algorithms such as Multinomial Na\"{\i}ve Bayes, Bernoulli Na\"{\i}ve Bayes, Logistic Regression, SGD classifier, SVC, Linear SVC and NuSVC for text classifications and number of algorithms like LexRank, Random Indexing, Latent Semantic, TextRank and Bm25 for text summarization.  In the final analysis we have found that SGD classifier has provided promising result during text classification after ensembling of text. We have got $95.8$\% accuracy by our selected approach on the novel dataset.

%
%

\end{document}